# QCD at finite baryon density with t-asymmetric fermions *

E. Mendel, L. Polley [a]

[a]FB Physik, Universität Oldenburg,
D-26111 Oldenburg, Germany

Susskind's continuous-time fermions, with two flavours, can be latticized using a one-sided time derivative. We are presently investigating the interacting case, where we hope to find the onset at finite $\mu$ at the right place due to the reduced number of flavours. As for these fermions there is only a discrete chiral symmetry left over, the lightness of pions in the broken phase has to be investigated.

## 1. INTRODUCTION

Based on the idea that the early onset of Baryon density as function of chemical potential, $\mu$, could be due to the excessive number of valence-quark flavours present in previous simulations, we have started to investigate *t-asymmetric fermions*, where we can reduce to two flavours.

In the past, several [1–3] lattice QCD simulations at finite $\mu$ had shown an onset in thermodynamic quantities at a $\mu$ much lower than at the expected $m_N/3$. One of us had proposed [4,2] that this early onset could be due to the fact that we had been using the usual Kogut-Susskind fermions, thus having a 4-flavour valence-quark degeneracy, in contrast to the 2 light flavours present in nature. The lattice fermions can change their flavour content along the propagation in the interacting case, producing in this way many more light nucleon states than in nature, thus weakening the Fermi repulsion, while on the other hand increasing the scalar Yukawa attraction due to the extra light scalar channels. Both effects would increase substantially the nuclear binding, producing the observed early onset on the lattice.

In order to reduce the quark flavour degeneracy we first suggested the use of Wilson fermions, where one has just 1 flavour, but they are harder to implement due to the Dirac components besides their explicit breaking of chiral symmetry. That project is being done with an innovative *volume method* to obtain the fermionic expectation values and its first results are reported in this proceedings [5].

Here we would like to present another fermion formalism, which reduces to 2 the flavour number [6], obtained by using a t-asymmetric fermionic action (one-sided time derivative). This action is unitarily equivalent to Susskind's Hamiltonian formalism [7,8], but in a simpler notation. The Heidelberg collaboration [9] were the first to investigate a finite $\mu$ for these fermions, in a version with 4-component spinors.

We have shown that at least in the free case [6] these fermions behave as 2 genuine flavours. We have constructed their Fock space and proven that in the continuum limit they will contain particles and antiparticles with the usual spectrum for the Hamiltonian and the Baryon number operators. Furthermore, the transfer matrix can be made positive, in contrast to the usual staggered fermions case.

In this work, we will introduce the formalism in the $SU(3)$ interactive case and present the first simulation results.

## 2. t-ASYMMETRIC FERMIONS

Usually we discretize the fermionic action, $S_F$, with a symmetrical derivative

$$\partial_\nu \psi = \frac{1}{2a}\left(\psi(x+\hat\nu) - \psi(x-\hat\nu)\right)$$

in order to get good hermiticity properties for $S_F$. This produces the flavour doubling in each coordinate direction. But in order to have a sensical theory we only need an *hermitian hamiltonian*

*Talk presented by E. Mendel

(eventually, on the lattice, a positive transfer matrix). This means that we don't have to take a symmetrical time derivative, thus reducing by 2 the flavour number.

We choose then an action $S_F = \bar\psi_x K_{xy} \psi_y$, where:

$$K_{xy} = m\delta_{xy} + \frac{1}{a_\tau}\,\gamma_4 \left( e^{\mu}\, U_{x,4}\, \delta_{x,y-\hat{4}} - \delta_{x,y} \right) \quad (1)$$

$$+ \frac{1}{2a_\sigma} \sum_k \gamma_k \left( U_{x,k}\, \delta_{x,y-\hat{k}} - U^{\dagger}_{x-\hat{k},k}\, \delta_{x,y+\hat{k}} \right)$$

To reduce further the flavour degeneracy, we can absorb a $\gamma_4$ into the $\bar\psi$ functional integration and then, diagonalizing spinors in analogy to Kawamoto-Smit:

$$\psi(x) = \alpha_1^{x_1} \alpha_2^{x_2} \alpha_3^{x_3} \chi(x) \quad \text{with} \quad \alpha_k = i\gamma_4\gamma_k \qquad (2)$$

we get,

$$K'_{xy} = m\gamma_4 \Gamma_4(x) \delta_{xy} + \frac{1}{a_\tau} \left( e^{\mu} U_{x,4}\, \delta_{x,y-\hat{4}} - \delta_{x,y} \right)$$

$$- \frac{i}{2a_\sigma} \sum_k \Gamma_k(x) \left( U_{x,k}\, \delta_{x,y-\hat{k}} - U^{\dagger}_{x-\hat{k},k}\, \delta_{x,y+\hat{k}} \right) (3)$$

with the usual: $\Gamma_\nu(x) = (-)^{x_1+\cdots+x_{\nu-1}}$. Taking just the first component of this diagonal kernel $K'$ (like the thinning out for K.S. fermions) we have managed to reduce to just 2 quark flavours. Note that our fermions are unitarily equivalent to Susskind's hamiltonian formalism ones [7], by: $\chi_{\rm our} = i^{x+y+z} (-1)^{x\cdot z} \chi_{\rm their}$.

Susskind showed [7] that there is a *discrete* version of chiral invariance left over, preventing renormalization mass counterterms in the interacting case. This is a nice feature as we know then that the chiral limit in the continuum limit should be at $m_q = 0$, in contrast to Wilson fermions. Unfortunately, although the chiral condensate seems to be spontaneously broken from our first simulations (also at strong coupling [8]), as the chiral symmetry is only discrete, there is no guarantee that the $\pi$ will be a Goldstone boson. We see that these fermions have some "intermediate behaviour" in its properties, in between usual Kogut-Susskind and Wilson ones.

Even though there is no usual Goldstone mechanism, we could still get light pions, in a similar fashion to Aoki's proposal [10] for Wilson fermions: Close to a second-order phase transition, to a phase where the order parameter (for parity), $\langle\pi\rangle$, gets an expectation value, the correlation length for $\langle\pi(0)\pi(t)\rangle$ diverges, giving a massles pion. In the Wilson case one can shift $\kappa$ in order to get close to the chiral limit, but here we have to get the critical point at $m_q = 0$, which will presumably only be reached in the continuum limit.

A potential problem with these t-asymm. fermions, in contrast to the usual formulations, lies in the fact that the action $S_F$ is *not* manifestly $O(4)$ invariant, endangering relativistic covariance. One could even naively ask from a hopping parameter expansion, if there is backward propagation at all! In fact, for free fermions we have checked that the antiperiodic boundary condition plays a crucial role in making positive-energy modes propagate forwards and negative backwards, in a symmetrical way in the continuum limit (for modes far from the cutoff). It will be important to check that this also happens when calculating a meson propagator as it consists of a particle and an antiparticle propagator.

Furthermore, even if the discretization in the action looks naively symmetric for spatial and time directions (when $a_\sigma = a_\tau$), the fermion propagator has to be renormalized (even in the quenched case by gluon loops) which will almost certainly destroy the $O(4)$ symmetry. Assuming that the action is renormalizable (which is as yet not clear), we would get renormalization constants $Z_\sigma$ and $Z_\tau$ in front of the spatial and temporal derivative terms in $S_F$. One can absorb $\frac{Z_\tau}{a_\tau}$ in the $\chi$ fields but then one gets a factor $\frac{Z_\sigma}{Z_\tau}\frac{a_\tau}{a_\sigma}$ in front of the spatial term. Thus, in order to regain $O(4)$ invariance we could tune the parameter $\xi \equiv \frac{a_\sigma}{a_\tau}$, as if we were on naively asymmetric lattices, until we regain the symmetry. This could be controlled non-perturbatively by calculating some physical propagator in several directions or by evaluating for some states the dispersion relation:

$$E a_\tau = \sqrt{\sum_j \left(\frac{a_\tau}{a_\sigma}\right)^2 \sin^2(p_j a_\sigma) + (m a_\tau)^2} \qquad (4)$$

so as to approach the relativistic one for some $\xi$.

## 3. STATUS OF COMPUTATIONS

We have already done first simulations of the baryon density, $\langle J_4 \rangle = \langle e^{\mu}\, \bar{\chi}_x\, U_{x,4}\, \chi_{x+\hat{4}} \rangle$ and the chiral condensate, $\langle \bar{\chi}\chi \rangle$, as function of $\mu$. The operators have been calculated with the solved pseudofermion method [2]. The curves for $\langle J \rangle$ are shown in fig. 1 and show a clear onset for both $m_q$ for $\mu \approx 0.2$. This is in clear contrast with our results for staggered fermions where there was a clear shift among the onsets for both masses. If the onset would be controlled by $m_\pi/2$ (as argued by some authors [11]), *and if the pion is a Goldstone boson for these fermions*, then the onset should be at twice $\mu$ for the higher mass. This shows that it is crucial to calculate the hadron masses in order to decide if the onset is at $m_N/3$ as wanted, or at some lower value due to some non-Goldstone meson. The vacuum value [6] of $\langle J \rangle = 1.5$ corresponds to saturation of the Fermi sea for these fermions ( $2_f \times 2_s \times 3_c$ per cube of size $2^3$).

The hadron masses can be calculated from the decay of the corresponding zero-momentum correlation of operators [8]. For the $\pi$:

$$\pi_0(r) = \chi_r^\dagger\, U_{r,z}\, \chi_{r+\hat{z}} + \chi_{r+\hat{z}}^\dagger\, U_{r+\hat{z},z}^\dagger\, \chi_r \tag{5}$$

and for the neutron with spin down, e.g.,

$$n_\downarrow(r) = \tfrac{1}{4}\left(1 - (-)^z + (-)^{x+y} + (-)^{x+y+z}\right) \times \tag{6}$$
$$[1 - i + (-)^y (1+i)]\, (-)^{x+z}\, \epsilon_{ABC}\, \chi_r^A \chi_r^B \chi_r^C$$

We are presently starting to calculate these masses and studying the $O(4)$ symmetry on improved algorithms in order to get the necessary precisions.

## 4. CONCLUSIONS

We have seen that for both $m_q$'s the onset is at very similar $\mu$'s indicating that it could be controlled by $m_N/3$ as wanted. It could also be that the $\pi$ is not Goldstone-like for these fermions, thus being able to give so similar onsets. We are currently investigating the masses and renormalization effects in order to decide this question. We would like to thank the RRZN in Hannover for supercomputer time.

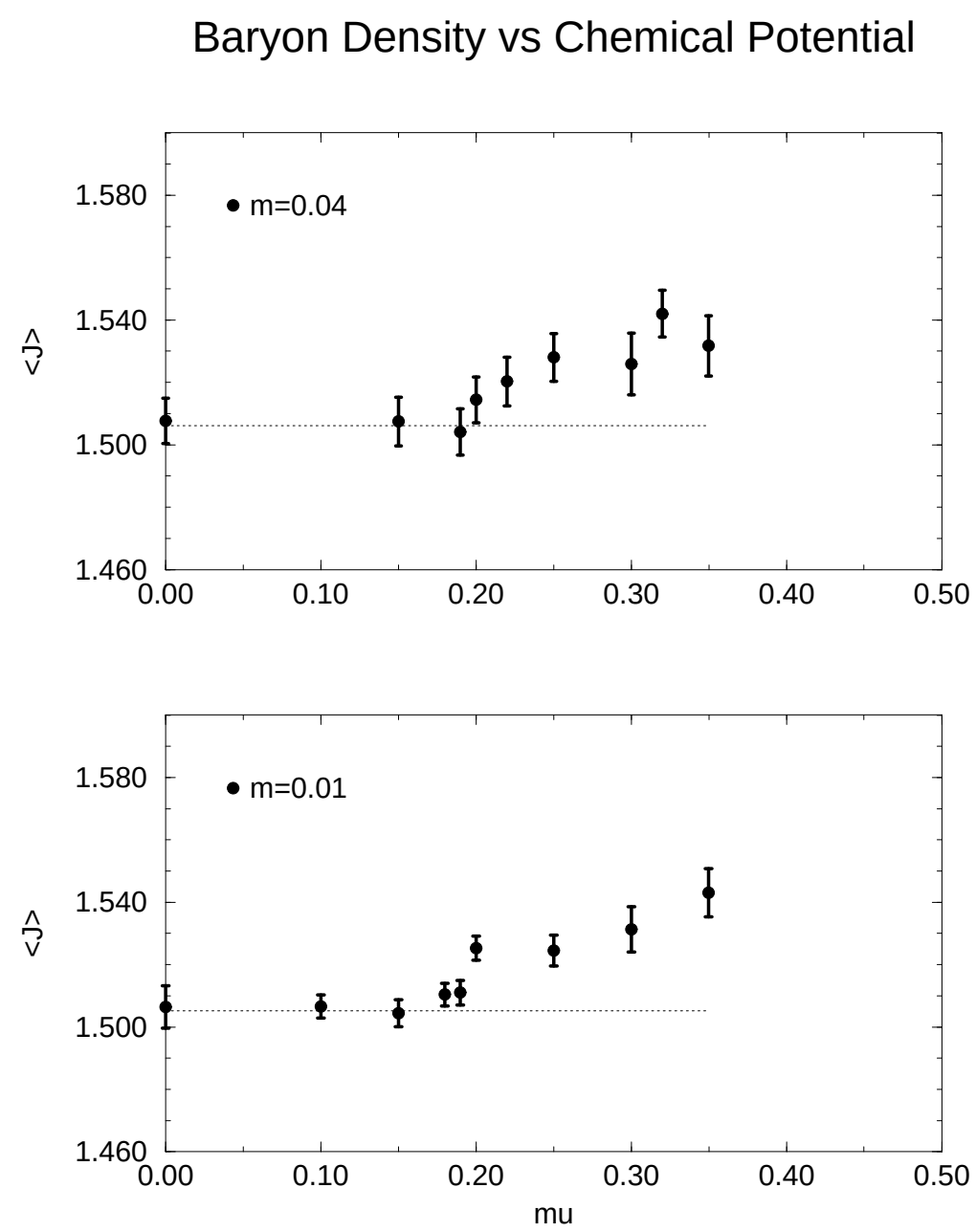

Figure 1. $\langle J \rangle$ vs $\mu$ for two $m_q$. The nearby onset is inconsistent with a Goldstone-like $m_\pi/2$.